\documentclass[a4,11pt]{article}
\usepackage{epsfig}

\usepackage{amssymb,epsfig,graphicx,epsf,epstopdf}

\usepackage[titletoc,toc,title]{appendix}

\usepackage{amsfonts}
\usepackage{amsmath}
\usepackage{feynmf}

\usepackage{latexsym}
\usepackage{float} 

\usepackage{slashed}
\usepackage{cite}
\usepackage[utf8]{inputenc} 
\usepackage{mathtools}
\usepackage[font=small]{caption}
\usepackage{color}
\DeclareFontFamily{OT1}{pzc}{}
\DeclareFontShape{OT1}{pzc}{m}{it}{<-> s * [1.10] pzcmi7t}{}
\DeclareMathAlphabet{\mathpzc}{OT1}{pzc}{m}{it}

\usepackage{commath} 
\usepackage{accents} 
\usepackage{relsize} 
\usepackage[makeroom]{cancel} 
\usepackage{xcolor}
\usepackage[normalem]{ulem} 

\usepackage[colorlinks=true, allcolors=black]{hyperref} 

\newcommand{\ben}{\begin{equation}}
	\newcommand{\een}{\end{equation}}
\newcommand{\be}{\begin{equation*}}	
	\newcommand{\ee}{\end{equation*}}

\newcommand{\ba}{\begin{eqnarray}}
\newcommand{\ea}{\end{eqnarray}}
\newcommand{\bal}{\begin{aligned}}
\newcommand{\eal}{\end{aligned}}

\begin{document}
\title{Measuring $H_0$ with pulsar timing arrays}

\author{
Dom\`enec Espriu, Luciano Gabbanelli and Marc Rodoreda\\[3ex]
{\it Department of Quantum Physics and Astrophysics and}\\ 
{\it Institut de Ci\`encies del Cosmos (ICCUB), Universitat de Barcelona,}\\
{\it Mart\'\i ~i Franqu\`es 1, 08028 Barcelona, Spain.}}

\date{}

\maketitle

\begin{abstract}
Pulsar Timing Arrays have yet to convincingly observe gravitational waves. Some time ago it was pointed out by one of the authors that a dramatic enhancement of the signal would take place for particular values of the angle subtended by the source and the observed pulsar. This enhancement is due to the fact that waves propagate in a Friedmann-Lemaitre-Robertson-Walker metric where, contrary to some wide-spread belief, a simple harmonic function with a red-shifted frequency is not a solution of the equation of motion. At the first non-trivial order, proper solutions have an effective wave number that differs from the frequency. This leads to some interesting effects in Pulsar Timing Arrays whose most visible manifestation is the enhancement of the signal that, all other parameters kept fixed, is related in a simple manner to the value of $H_0$. In this work, we rederive in an alternative way the main results, extend the formalism to a more realistic setting where all components in the cosmological budget are included, investigate in detail the dependence of the signal on the various parameters involved and propose an observational set-up to hopefully detect this very relevant effect.
\end{abstract}
\vfill
\noindent

\noindent
ICCUB-19-014

\section{Introduction}
The Hubble constant $H_0$ describes the current rate of expansion of the Universe and contains information of its composition.
In the last decades, technological progress has produced a qualitative change in observations, decreasing significantly 
the error bars and making $H_0$ one of the better measured cosmological parameters. Yet, at present, there is a 
controversy about its actual value: a significant incompatibility between local and non-local measurements 
of the Hubble constant has been inferred from recent data.  There is an apparent tension of more than $4\sigma$ and 
less than $6\sigma$ \cite{Verde} between the measurement of $H_0$ in the late Universe using standard candles \cite{Riess} 
and in the early Universe with the CMB spectrum \cite{H0Plank}. For this reason, there is  great interest in 
achieving new and independent measurements of this cosmological quantity.

An alternative method to measure the Hubble constant is standard sirens, the analogy of an astronomical standard candle using gravitational waves. This possibility remained on the waiting list until the first detection of gravitational waves was announced by LIGO and Virgo collaborations \cite{LIGOFirst}.  On August 17th 2017, the gravitational-wave event GW170817 and the gamma-ray burst GRB 170817A were observed independently, corresponding to a binary neutron star merger and the first multi-messenger observation \cite{L12}. This event was used as a standard siren and led to the first-ever measurement of the Hubble constant using gravitational waves \cite{NatureFirstH0}. Later, another measurement of $H_0$ was done, this time with a binary black hole merger as a standard siren, combined with a photometric redshift catalog from the Dark Energy Survey (DES) \cite{L7}.

Other relevant gravitational-wave experiments are Pulsar Timing Arrays (PTA), although a direct detection of gravitational 
waves using this strategy has yet to be achieved. PTA stand for a set of millisecond pulsars as well as for the collaborations that carefully 
monitor them in order to detect anomalies that might signal the passage of a gravitational wave through our 
galaxy \cite{Romani1989FosterBecker1990}. PTA exploit the remarkable period stability of millisecond pulsars and explore 
correlated timing variations among the pulsars in the array. 

Three different collaborations have been set up in three different places and have been taking data for some time. 
Specifically, the Australian Parkes PTA (PPTA) \cite{Manchester2013}, the North-American Nanohertz Observatory for 
gravitational waves (NANOGrav) \cite{Arzoumanian2015} and the European PTA (EPTA) \cite{Desvignes2016}. 
Combined, they form the International PTA (IPTA) as described in \cite{Hobbs2010ManchesterIPTA2013}. 
This collaboration monitors around 50 pulsars in the quest for finding footprints of the passage of gravitational 
waves in the aforementioned correlated signals.

Conventionally, gravitational waves are treated as consisting of a superposition of freely propagating waves. This stems
from the fact that our spacetime is certainly almost flat, i.e. not very different from Minkowski spacetime 
at short distances. In the flat spacetime case, considering small perturbations around the Minkowski metric, linearization of the Einstein field equations leads to a homogeneous wave equation, corresponding to gravitational waves 
perturbations. In the familiar transverse and traceless (TT) gauge the equation for the perturbation $h_{\mu\nu}$ 
in $g_{\mu\nu}=\eta_{\mu\nu}+ h_{\mu\nu}$ takes an exceedingly simple form. In vacuum it is just
$\Box h_{\mu\nu} =0 $, whose solution consists in a combination of harmonic functions. The gauge condition imposes some
restriction on the form of $h_{\mu\nu}$ that will be discussed later.
The effect of being in an expanding universe is taken into account by considering a red-shifted set of frequencies.

In \cite{Espriu1,Espriu2} it was seen that this issue needed further analysis. The coordinates that are natural where and when the
gravitational waves are produced are not the ones used when detecting the gravitational waves. The change of coordinates
(worked out in \cite{Espriu3} when all cosmological components are included) implies not only the pertinent red-shift in 
the frequency of the waves, but also a different modification of the wave number. This modification does not
affect local interferometric experiments (such as LIGO or Virgo \cite{LIGOVirgo}) but is relevant in the case where the distortion
induced by the wave has to be integrated over a sufficiently long distance. This effect was seen to be highly relevant
for the IPTA project observations, to the point that it could actually facilitate the first detection of gravitational waves
in such a setting.

In this work we provide a completely different derivation of the physical origin of the effect. We also analyse in detail the
dependence of the signal on various parameters: frequency, varying amplitude, distance to the source and to the pulsar.
We conclude with a discussion on the possibility that the effect discussed in this work could facilitate a first detection 
of gravitational waves in PTA and could also provide an alternative method for measuring the Hubble constant $H_0$.

\section{Linearized perturbations in FLRW}
At cosmological distances, the propagation of gravitational waves takes place not over 
a flat spacetime, but over an approximately globally de Sitter universe, described by the Friedmann–Lemaitre–Robertson–Walker (FLRW) metric. Applying the strategy of linearization on a 
curved background, we consider small perturbations around the FLRW background metric\footnote{We
denote FLRW coordinates always by capital letters.} 
\begin{equation}
ds^2= \tilde{g}_{\mu \nu} dX^\mu dX^\nu = dT^2 - a^2(T) (dR^2 + R^2d\Omega^2)\ ,
\end{equation}
of the form
\begin{equation}
g_{\mu \nu} = \tilde{g}_{\mu \nu} + h_{\mu \nu}, \qquad |h_{\mu \nu}| \ll 1\ .
\end{equation}
These perturbations are interpreted as gravitational waves propagating on this curved background in 
analogy to the treatment on a flat spacetime. At first order, the inclusion of this perturbation on the set 
of Einstein equations gives the following expansion
\begin{equation}
G_{\mu\nu}(\tilde{g}+h) = G_{\mu\nu}(\tilde{g}) + 
\frac{\delta G_{\mu\nu}}{\delta g_{\alpha \beta}}\, \biggl\rvert_{\tilde{g}}\ h_{\alpha\beta} + \ .\ .\ .\ 
= \kappa \left(T_{\mu\nu}^{\scriptscriptstyle\,(0)} + T_{\mu\nu}^{\scriptscriptstyle\,(1)}+ \ .\ .\ .\ \right)
\end{equation}
where $\kappa=8\pi G / c^4$, $G_{\mu \nu}$ is the Einstein tensor and $T_{\mu \nu}$ is the energy-momentum tensor. 
Clearly, Einstein's equations are satisfied for the unperturbed background FLRW 
metric, $G_{\mu\nu}(\tilde{g})=\kappa T_{\mu\nu}^{\scriptscriptstyle\,(0)}$, which in fact gives the well-known Friedmann 
equations by treating the energy content of the universe as perfect fluids. The energy-momentum 
tensor includes all components of the cosmological background, each one being 
described by some equation of state that relates the pressure and the density for each component, 
$P_i= \omega_i \rho_i$. The cosmological constant is included in the energy-momentum tensor as
a constant density fluid \cite{Peebles} with $ \omega_{ \scriptscriptstyle \Lambda}= -1$. Hence
\begin{equation}
T_{\mu\nu}^{\scriptscriptstyle (\Lambda)}=\rho_{ \scriptscriptstyle \Lambda}\,g_{\mu\nu}\ ,
\end{equation}
where $\rho_{\scriptscriptstyle\Lambda}=-p_{\scriptscriptstyle\Lambda}=\Lambda/\kappa$.
By Gauss outflux theorem, the cosmological constant is equivalent to a repulsive
gravitational field of magnitude $\Lambda\,r/3$ away from any chosen center. The current preferred value 
is $\Lambda= 1.11 ~10^{-52}$ m$^{-2}$ \cite{LambdaValue}. The other components are
\begin{equation}
T_{\mu\nu}^{\scriptscriptstyle (dust)}= \rho_{\scriptscriptstyle dust} \,  U_{\mu} U_{\nu}\  ,\qquad T_{\mu\nu}^{\scriptscriptstyle (rad)}= \frac{4}{3}\rho_{\scriptscriptstyle rad} \,  U_{\mu} U_{\nu} - \frac{1}{3}\rho_{\scriptscriptstyle rad} \,  g_{\mu \nu}\ , 
\end{equation}
since $P=0$ for dust and $P=\rho / 3$ for radiation. The four-velocity fulfills the normalization 
condition $g_{\mu \nu} \, U^{\mu} U^{\nu} = 1$.
Then, at first order, we end up with the following expression, involving the perturbed Einstein tensor
\begin{equation}\label{Eq: Final Perturbations}
\frac{\delta G_{\mu\nu}}{\delta g_{\alpha \beta}}\, \biggl\rvert_{\tilde{g}}\ h_{\alpha\beta} = 
\kappa \, T_{\mu\nu}^{\scriptscriptstyle\,(1)}\ .
\end{equation}
Evaluating the left hand side of Einstein's equations requires the computation of the Ricci tensor and 
its contracted form, the Ricci scalar, for the perturbed metric, which are functions of the Christoffel 
symbols. In essence, the procedure is the same than doing perturbations around the flat Minkowski metric, 
but in this case there are many more terms. Before proceeding, in order to avoid redundancies under 
coordinate transformations, we must choose a gauge. It is convenient to work in the transverse and traceless (TT) 
gauge, where this set of equations are simplified. Also, it is reasonable to expect that 
in spherical coordinates gravitational waves propagate along the radial direction away from the source
along a vector given by $k^\mu=(w,k_{\scriptscriptstyle R},0,0)$. Then, transversality implies that the 
only non-vanishing $h$-components are the purely angular ones, i.e.
\begin{equation}
h_{T \alpha} = h_{R \alpha} = 0
\end{equation}
for $\alpha = \{T,R,\theta, \phi \}$. Moreover, the traceless condition relates the diagonal 
components of the metric perturbation by
\begin{equation}
h = \tilde{g}^{\mu \nu} \, h_{\mu \nu} = 0 \quad \Rightarrow \quad h_{\phi \phi} = - h_{\theta \theta} \, \sin^2 \theta\ .
\end{equation}
Together with $h_{\theta \phi} = h_{\phi \theta}$ due to the symmetry of the metric tensor, there are 
two non-zero independent components, in agreement with the two degrees of freedom of gravitational waves. 

The two equations for $h_{\theta \phi}$ and $h_{\phi \phi}$ can be obtained from the 
perturbed Einstein's equations \eqref{Eq: Final Perturbations} after imposing the gauge condition. 
Up to first order in the perturbations one gets (for simplicity we do not depict the dependence
of the cosmological scale factor on $T$)
\begin{equation}\label{Eq: Wave Eq FLRW}
\left[\frac{a}{2}\ \frac{d}{dT}\left(\frac{1}{a}\,\frac{d}{dT} \right) - 
\frac{R^2}{2a^2}\frac{d}{dR}\left(\frac{1}{R^2}\,\frac{d}{dR}\right) - 
\frac{1}{R^2\,a^2} - \left(\frac{\dot{a}}{a}\right)^2 - \frac{3\,\ddot{a}}{a}\right]\ h_{\mu\nu} 
= \kappa \, T_{\mu\nu}^{\scriptscriptstyle \,(1)}\ ,
\end{equation}
where $\{\mu,\nu\}=\{ \theta,\phi\}$.
First of all, we notice that the equations satisfied by the four non-vanishing metric components have exactly 
the same functional form and no angular dependence. The above equation would be analogous to $\Box h_{\mu\nu} =0 $, fulfilled by the linearized metric perturbations in Minkowski spacetime.

\begin{figure}[t!]
	\centering
	\includegraphics[width=0.45\linewidth]{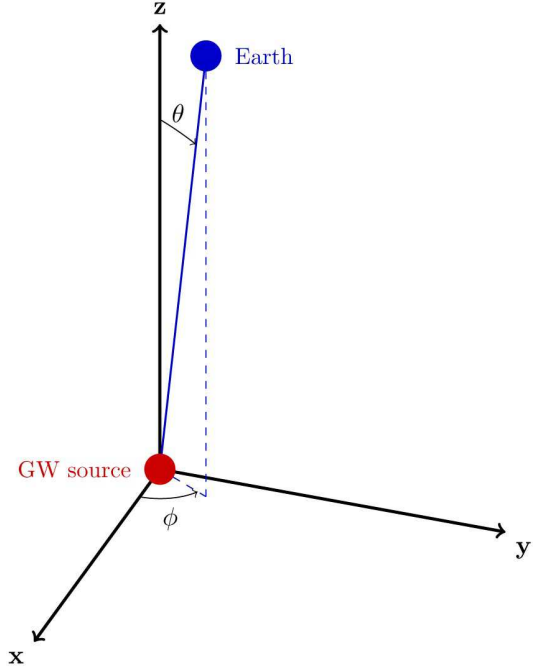}
	\caption{Diagram of the situation we are considering, with the usual definition of spherical coordinates with angles $\theta$ and $\phi$. 
	Since the Earth is very distant from the gravitational wave source, we are interested in the $\theta \approx 0$ case.}
	\label{Fig: Spherical coord}\end{figure}

Since we observe gravitational waves emitted by a very distant source, we are very interested in the concept of 
plane wave solutions. A more useful coordinate system for doing this analysis is the Cartesian set of coordinates. We can always choose spatial coordinates to make the gravitational wave travelling in the Z direction with the source located in the X--Y plane, as shown 
in Figure \ref{Fig: Spherical coord}.
In the transverse and traceless gauge, for a wave propagating with wave 
vector $k^{\mu} = (\omega, 0, 0, k^3)$, transversality implies
\begin{equation}
h_{T \alpha} = h_{Z \alpha} = 0\ ,
\end{equation}
so the non-zero components are $h_{XX}$, $h_{YY}$, $h_{XY}$ and $h_{YX}$. In these coordinates, gravitational waves 
coming from a very distant source correspond to a very small polar angle case, $\theta \approx 0$. After applying 
the transformation in this polar angle limit, we end up with the following relations 
\begin{eqnarray}
&h_{XX} &=\hspace{8pt}  \frac{1}{R^2} \cos(2\phi) \, h_{\theta \theta} - \frac{1}{R^2} \frac{\cos \theta}{\sin \theta} \sin(2\phi) \, h_{\theta \phi}\\[2ex]
&h_{YY} &= -\frac{1}{R^2} \cos(2\phi) \, h_{\theta \theta} + \frac{1}{R^2} \frac{\cos \theta}{\sin \theta} \sin(2\phi) \, h_{\theta \phi}\\[2ex]
&h_{XY} &=\hspace{8pt}  \frac{1}{R^2} \sin(2\phi) \, h_{\theta \theta} + \frac{1}{R^2} \frac{\cos \theta}{\sin \theta} \cos(2\phi) \, h_{\theta \phi}\, .
\end{eqnarray}
Note that the apparent singularity in the prefactor of $h_{\theta \phi}$ has to be compensated by the appropriate
vanishing of this component.

Before proceeding further, we can make a few comments. We notice that these relations satisfy 
$h_{XX} = - h_{YY}$ and $h_{XY} = h_{YX}$, which is to be expected since we are working with a symmetric tensor propagating 
on a background metric and in the transverse and traceless gauge. Also, we recover the familiar degrees of freedom, the
``plus" $h_{+}$ and ``cross" $h_{\times}$ polarizations. A passing gravitational wave in the Z--direction only disturbs 
test particles in directions perpendicular to the wave vector, similar to the electromagnetic case \cite{Carroll}. 
Moreover, we observe that under a rotation of the azimuthal angle from $\phi = 0$º to $45$º on the X--Y plane, the $h_{+}$ 
polarization becomes $h_{\times}$ and vice versa. This effect is a consequence of the azimuthal symmetry respect 
to the propagation Z--direction and we recover the known result that the two polarizations follow the same pattern 
but rotated $45$º with respect to the X--Y plane. We also notice that the polarization modes are invariant 
under rotations of $\phi = 180$º in the X--Y plane, as it is expected for spin-2 particles such as gravitons.

By direct substitution of the relations between $h_{XX}$, $h_{YY}$, $h_{XY}$ and $h_{\theta \theta}$, $h_{\phi \phi}$, $h_{\theta \phi}$ 
in the small angle limit $\theta \approx 0$, we can easily derive the corresponding Eq. \eqref{Eq: Wave Eq FLRW} 
in Cartesian coordinates. Before doing that, let us first make some approximations in the equation. 
At this point we are interested in keeping linear terms in the Hubble constant $H_0$ only.  
Recall that
\begin{equation}\label{HubbleConstant}
H_0 = \sqrt{\frac{\Lambda}{3} + \frac{\kappa \rho_{d0}}{3}+ \frac{\kappa \rho_{r0}}{3}}\ ,
\end{equation}
where $\rho_{d0}$ and $\rho_{r0}$ are the current energy densities of dust and radiation, respectively. 

Although Eq. \eqref{Eq: Wave Eq FLRW} is valid up to higher orders in  $\Lambda$, we will be interested in terms of order ${\cal O}(\sqrt{\Lambda})$ (ditto for the other components in $H_0$). Hence, in the 
right hand side of the aforementioned equation, we can get rid of the term $\kappa \, T_{\mu\nu}^{\scriptscriptstyle \,(1)}$ 
because it is linear with the densities and therefore of order ${\cal O}({H_0^2})$, as well as the latter two terms of 
Eq. \eqref{Eq: Wave Eq FLRW}, proportional to the square of the first time derivative of the scale factor 
and to the second time derivative. Then, the wave equation for the Cartesian metric perturbations
but written in spherical coordinates reads
\begin{equation}\label{Eq: Cartesian wave equation}
\left[\partial^2_T - H_0 \, \partial_T - \frac{1}{a^2}\left(\partial^2_R + \frac{2}{R}\,\partial_R \right)\right] h_{ij} = 0
\end{equation}
where now $i,j = {X,Y}$, $H(T) \equiv \dot{a}/a$ is the Hubble parameter and we have introduced the $\partial_i$ 
notation for derivatives. As before, the functional form of this equation is valid for the four non-vanishing 
metric perturbations in the transverse gauge $h_{XX}$, $h_{YY}$, $h_{XY}$ and $h_{YX}$. 
In the limit of an empty universe, with no cosmological constant or dust, the scale factor $a(T)$ can be taken to be
$a_0 = 1$, so Eq. \eqref{Eq: Cartesian wave equation} reduces to a homogeneous wave equation $\Box_{\tilde{\eta}}h_{\mu\nu}=0$, 
as expected for the flat spacetime limit.

Restricting ourselves to terms of ${\cal O}(H_0)$ can be justified based on the smallness of $H_0$ (and of all its 
ingredients, of course), since we will be interested in sources at distance $R \ll H_0^{-1}$, and also to ease the
comparison with previous work. In practice, since there are strong numerical cancellations, this will restrict 
the validity of the approximations made to distances of $\sim 1$ Gpc.

\section{Relation to previous work}
Given that gravitational waves observed by laser interferometry and potentially observed in PTA come from binary mergers, 
we are interested in these types of systems. Being a central-force problem and due to the approximately spherical 
symmetry of the system, once we are well away from the source, this type of gravitational waves is well described by 
a superposition of harmonic functions periodic in time $t$ in  Schwarzschild--de Sitter (SdS) coordinates $\{t,r,\theta,\phi\}$.
In the TT gauge
\begin{equation}\label{Eq: harmonic function}
h_{ij}(w;t,r)= \frac{e_{ij}}{r}\,\cos\left[w\,(t-r)\right]\, ;
\end{equation}
where $e_{ij}$ is the traceless and transverse polarization tensor, and $w$ the angular frequency. 
These coordinates are different from FLRW ones.
If at long distances from the source the universe was approximately Minkowskian, this wave front would be seen 
by a remote observer with exactly the same functional form. However, as discussed in 
\cite{Espriu1,Espriu2,Espriu3,Espriu4}, this coordinate system is not useful in cosmology since cosmological 
observations are done in comoving coordinates $\{T,R\}$. This coordinate transformation can be expanded 
in powers of energy densities \cite{Espriu3}. At the leading order (but not beyond that) the corrections
can be expressed in terms of the Hubble constant $H_0$ \cite{Alfaro} as follows
\begin{equation}\label{ChVarCCLinear}\left\{\begin{split}
\hspace{7pt} t(T,R) &= T + \frac{R^2}{2} \, H_0 + \ldots 
\\
r(T,R) &= R \left( 1 + \Delta T H_0\right) + \ldots 
\end{split}\right.
\end{equation}
At the first non-trivial order, all cosmological densities appear in such a combination so as to 
reproduce the Hubble constant $H_0$. This is not so when higher order corrections are considered. In principle
by going to the next order one would be able to discriminate between the different densities.

As discussed before, we are interested in the first order modifications, which as we just said correspond to 
first order of the Hubble constant. After transforming to FLRW comoving coordinates, a harmonic function 
\eqref{Eq: harmonic function} in SdS coordinates will get modified to
\ben\label{gwfrw}
h_{ij} \simeq 
\frac{e_{ij}}{R}\left(1 + H_0\,T\right)\cos\left[w\,(T-R)+w\,H_0\left(\frac{R^2}{2}-TR\right)\right]\,.
\een
The polarization tensor will not be very relevant 
for the following discussion, although its precise form is essential for precise comparison with 
observations. Due to the coordinate transformation, we observe that some anharmonicities of order $H_0$ 
have appeared. This last expression can also be written as
\ben\label{gwfrw3}
h_{ij}=
\frac{e_{ij}}{R}\left( 1 + H_0\,T\right)\cos\left[w_{\scriptstyle eff}\,T-k_{\scriptstyle eff}\,R \right]\,,
\een
with $w_{\scriptstyle eff}= w\bigl(1-H_0\,R\bigr)$ and $k_{\scriptstyle eff}= w\bigl(1- H_0\,R/2\bigr)$.
Notice that $w_{\scriptstyle eff}$ agrees with the usual frequency redshift (in a linearized approximation
suitable for small redshifts $z=RH_0$), so 
this well-known result is well reproduced. However, the wave number is different. This discrepancy will be 
at the root of a possible measurement of $H_0$ in PTA, as we will see below (eventually of the
separate densities entering $H_0$ if next-to-leading corrections are considered). The corrections of 
order $H_0^2$ and beyond are subleading when the actual presumed value for the cosmological constant 
is considered and distances to the gravitational wave source of the magnitude considered here are involved; only $H_0$ 
terms really matter for the physical situation we are interested.

The phase velocity of wave propagation in such coordinates is  $v_{p}\sim 1-H_0\,T + \mathcal{O}(H_0^2)$ \cite{Espriu1}. 
On the other hand, with respect to the ruler distance travelled (computed with $g_{ij}$) the velocity is still $1$. 
In the Lorenz gauge the only spatial components of the metric that are different from zero are the $X,Y$ entries 
of $e_{\mu\nu}$. 

The alert reader has noticed at this point that if all the reasoning leading to 
Eq. \eqref{gwfrw3} is correct, this expression should satisfy the equation of motion obtained by applying 
perturbation theory on a FLRW background metric, i.e. Eq. \eqref{Eq: Cartesian wave equation}. And indeed it does
up to terms of ${\cal O}(H_0)$ included. On the contrary, a simple harmonic function 
(including the frequency redshift) does not fulfill this equation. This shows that the reasoning above is valid
and that the anharmonicities induced by the coordinate transformation between SdS and FLRW coordinates are 
essential to fulfill the equation. As mentioned before, in the empty universe limit, these equations both reduce 
to the homogeneous wave equation in flat spacetime. In conclusion, by studying the differential equations obeyed 
by a perturbation on a background FLRW metric and also by carefully doing a coordinate transformation between the 
reference frame where gravitational waves originate (e.g. a black hole merger) and the reference frame where waves are measured, we obtain the same gravitational wave solution up to order $H_0$.

In fact, nothing would prevent us from going to the next order, namely to order $\Lambda$ according to the 
Hubble constant, Eq. \eqref{HubbleConstant}. The primordial equation, Eq. (\ref{Eq: Wave Eq FLRW}),
contains terms that were neglected in the previous discussion. However, in order to make a comparison with
the other method, the latter should include corrections of $O(\Lambda)$ too. Namely, one should find solutions
not only with a $1/r$ potential, but include $\Lambda r^2$ corrections as well.

We emphasize once more that while at leading order the results can be expressed in terms 
of the Hubble constant $H_0$ regardless of their composition, at the next order this is not so and one could 
discriminate, in principle,  among the various components.

\section{Relevance for Pulsar Timing Array}
It is clear that one has to understand what are the consequences of having $k_{eff}
\neq w_{eff}$. First of all, it should be stated right away that for laser interferometric experiments of the LIGO
or Virgo type, there is no consequence whatsoever. Such experiments only observe the local variation of the signal with time. They are
only sensitive to $w_{eff}$. In fact, in a merger neither the amplitude nor the frequency are constant, but are typically
time dependent (later the relevance of this will be noticed). In any case, the wave number is not measurable
in a local experiment.

However, this is not the case in PTA. These observations involve some non-local contributions as the signal
is integrated along the line of sight joining the pulsar and the observer. The fact that $H_0$ in \eqref{gwfrw} 
is non-zero (and hence $k_{eff} \neq w_{eff}$) may produce observable effects in PTA.

\begin{figure}[!b]
	\centering
	\includegraphics[scale=0.5]{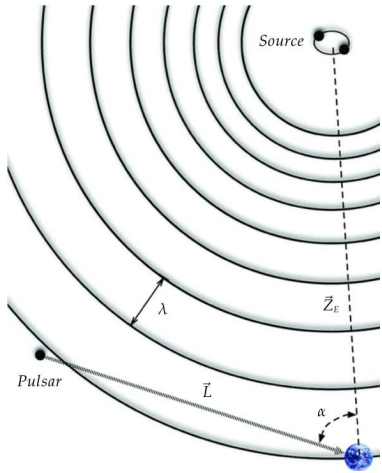}
	\caption{Relative coordinates of the gravitational wave source ($R=0$), the Earth (at $R=Z_E$ with respect to the source) and the pulsar (located at coordinates $\vec P= (P_X, P_Y, P_Z)$ referred to the source). 
	In FLRW coordinates centered on Earth, the angles $\alpha$ and $\beta$ are the polar and azimuthal angles of the pulsar with respect to the axis defined by the Earth-source direction.}
	\label{FigSetup}\end{figure}

Let us first review how this effect could be measured. Consider the situation shown 
in Figure \ref{FigSetup} describing the relative situation of a gravitational waves source (possibly a very massive 
black hole binary), the Earth and a nearby pulsar \cite{Deng}.
The pulsar emits pulses with an electromagnetic field phase $\phi_0$ . If this magnitude is measured from 
the Earth, it will be
\begin{equation}
	\phi(T)=\phi_0\left[T-\tfrac{L}{c}-\tau_0(T)-\tau_{\scriptscriptstyle GW}(T)\right]\ .\end{equation}
The corrections due to the spatial motion of the Earth within the Solar system and the Solar system with 
respect to the pulsar, together with the corrections when the electromagnetic wave propagates through the 
interstellar medium, are taken into account in $\tau_0$. The term $\tau_{\scriptscriptstyle GW}$ involves the 
corrections owing exclusively to the gravitational wave strength $h_{ij}$ computed 
in comoving coordinates. In the following we will focus in the phase corrections due to the passage of a gravitational wave \cite{Deng}; this shift is given by
\begin{equation}\label{TimingResidual}
	\tau_{\scriptscriptstyle GW}=-\frac{1}{2}\,\hat{n}^i\hat{n}^j\,{\cal H}_{ij}(T)\ .\end{equation}
The unit vector giving the direction can be written as 
$\hat{n}= (-\sin\alpha \cos\beta; -\sin\alpha \sin\beta; \cos\alpha)$, where
$\alpha$ is the angle subtended by the gravitational wave source and the pulsar as seen from
the Earth and $\beta$ is the azimuthal angle around the source-Earth axis. ${\cal H}_{ij}$  is the 
integral of the transverse-traceless metric perturbation along the null geodesic from the pulsar to the Earth. 
Let us choose the following parametrization: $\vec R(x)= \vec{P}+L(1+x)\hat{n}$ with $x\in[-1;0]$, $L$ 
the distance to the pulsar and $\vec{P}$ the pulsar location with respect to the source. Making the dependencies 
explicit, this magnitude is given by
\begin{equation}\label{tim}
	{\cal H}_{ij}=\frac{L}{c}\int_{-1}^0dx\ h_{ij} \bigl[\,T_E+\tfrac{L}{c}\,x\ ;\ \vec R(x)\,\bigr]\ .\end{equation}
$T_E$ is the time of arrival of the wave that travels a distance $Z_{E}$ to the local system. 
The speed of light has been restored in this formula. In the TT gauge, a gravitational wave propagating 
through the $Z$ axis has a polarization tensor with only nonzero values in the $X$, $Y$ components.

The distortion created by the intervening gravitational wave 
modifies the timing residual.  The real question is whether the fact that $k_{eff}\neq w_{eff}$ has
any unexpected consequences. This question was answered in the affirmative in \cite{Espriu1}. 
In order to see the magnitude of the effect we
take average values for the parameters involved: we assume for simplicity $\varepsilon \sim | e_{ij}|$, $i,j=X,Y$ 
(this approximation does  not affect in any significant way the results below, and in fact it can be easily removed)
and take the value $\varepsilon=1.2\times 10^9$ m for a gravitational 
wave generated by some source placed at $Z_{E}\sim (0.1 - 1)$ Gpc, giving a 
strength $|h|\sim \varepsilon/R\sim (10^{-16}-10^{-15})$ 
which is within the expected accuracy of PTA \cite{jenet}. Monitored pulsars are usually found at a distance 
of the order of $L\sim1$ kpc. These, together with the preferred value of $H_0$, are all the 
parameters we need to know.

On geometric grounds, we are always able to choose a plane containing the Earth, the pulsar and 
the source and set $\beta=0$. After all these considerations, in order to obtain the timing residual 
we need to perform the following integral according to previous results \cite{Espriu1,Espriu2,Alfaro}
\begin{equation}\label{TimingResidualH0}\begin{split}
	\tau_{\scriptscriptstyle GW}^{\scriptscriptstyle (\Lambda CDM)}= -&\frac{L\varepsilon}{2c}\ \sin^2 \alpha\ \int_{-1}^0\ dx\ \frac{1+H_0\ (\,T_E+x\,\tfrac{L}{c}\,)} {Z_E+x\,L\ \cos\alpha+\dots}\ \cos (w\,\Theta)\ ,\end{split}\end{equation}
where the argument of the trigonometrical function (up to a global phase) is given by
\begin{equation}\label{Theta}\begin{split}
	\Theta&\equiv x\ \tfrac{L}{c}\,\left(1-\cos\alpha\right)-H_0\ (\,T_E+x\, \tfrac{L}{c}\,\cos\alpha\,)\ \bigl[\,\tfrac{T_E}{2} +x\,\tfrac{L}{c}\, \left(1-\tfrac{\cos\alpha}{2}\right) \,\bigr]\, .\end{split}\end{equation}

\begin{figure}[!b]
	\centering\includegraphics[scale=1]{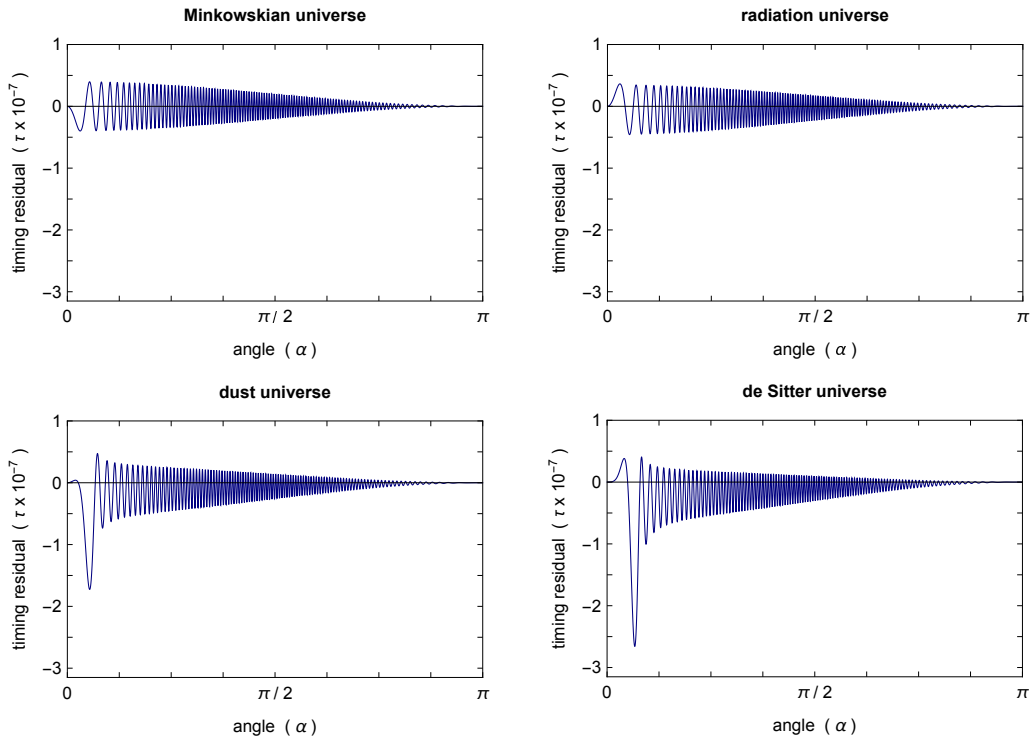}
	\caption{
		Raw timing residual as a function of the angle $\alpha$ subtended by the source and the measured pulsar as seen from the observer. The figures represent a snapshot of the timing residual at a given time $T_E$ in flat space time, in a universe with only relativistic matter, with only non-relativistic matter, and in a purely de Sitter universe. Clearly, at least in a superficial analysis, it seems plausible to be able to disentangle both contributions. The figures depict the signal at the time $T_E=Z_E/c$. Time evolution or small changes in the parameters will change the phase of the oscillatory signal but not the enhancement clearly visible at relatively low values of the angle $\alpha$. The figures are symmetrical for $\pi\leq\alpha\leq 2\pi$.}\label{FigTimingResidual1Comp}\end{figure}

Snapshots at $T=T_E=Z_E/c$ of the resulting timing residuals as a function of the angle $\alpha$ can 
be obtained for different values of the Hubble constant. 
In Figure \ref{FigTimingResidual1Comp} we compare these timing residuals for different universes characterized by various values of $H_0$. 
The distance to the source and the gravitational waves frequency are always chosen to be $Z_E=100$ Mpc 
and $w=10^{-8}$ Hz respectively, except when specifically indicated in the following plots. The figure 
speaks by itself and it strongly suggests that the angular dependence of the timing residual is influenced 
by the value of $H_0$, irrespectively of whether this value is due to a cosmological constant, 
radiation and/or dust. As explained below, at the end of this section,  at this order there is a degeneracy and it is the overall value of $H_0$ what
matters. The feature that catches the eye immediately is the enhancement of the signal for a specific 
small angle $\alpha$, corresponding generally to a source of low galactic latitude or a pulsar nearly 
aligned with the source (but not quite as otherwise $e_{ij}\hat n^{i}\hat n^{j}=0$). This enhancement
is not present if $H_0=0$.

Surprising as it is, this enhancement is relatively easy to understand after a careful analysis of the integral 
in Eq. \eqref{tim}. As was explained in \cite{Espriu1}, an approximate formula relating the angular location of the 
enhancement $\alpha$ to the value of $\Lambda$ can be derived (the effect is independent of the angle $\beta$). 
This relation can be more conveniently expressed in terms of the Hubble parameter $H_0$; in fact
\begin{equation}\label{RelationAngleHubble}
	H_0\simeq\frac{2c}{Z_E}\ \sin^2\frac{\alpha}{2}\end{equation}
as it has been shown in \cite{Alfaro}. This approximate relation is extremely interesting in two ways: on the one side, this prediction could be eventually 
tested experimentally and may facilitate enormously the detection of gravitational waves produced in supermassive 
black hole mergers. On the other side, and the most interesting for us, this relation may provide an indirect way 
to measure the Hubble constant, and provide a local ($z=Z_E H_0 \ll 1$) detection of the cosmological constant once the distance 
to a supermassive black hole binary $Z_E$ and the angular location of the enhancement $\alpha_{max}$ are known. 
It is particularly interesting to measure $H_0$ as locally as possible because its main component $\sqrt{\Lambda/3}$ 
is assumed to be an intrinsic property of the spacetime, present to all scales, something that needs to be proven.

The effect discussed in the previous sections is largely degenerate with respect to the details of the source 
of gravitational waves. Not much depends on the polarization or frequency of the wave front (see e.g. \cite{Espriu1}). 
This is so because the enhancement is entirely a consequence of the characteristics of the FLRW metric. 
The enhancement for a particular value of the angle subtended by the source of gravitational waves and the pulsar 
cannot be mimicked by changing frequency or amplitude of a gravitational wave, or polarization. All these changes 
would enhance or suppress the overall strength of the signal for all angles, not a restricted range of values. 
The phenomenon described is thus indeed a telltale signal. 

However, the figures seen above are only of theoretical
interest, to prove that the effect exists. What one needs to do is to define an observational protocol.

The observable defined in \eqref{TimingResidual} is, in fact, not necessarily the preferred one in PTA collaborations. 
Typically, PTA use two-pulsars correlator together with the assumption that the gravitational wave signal at the location of the two pulsars is uncorrelated (see \cite{Hellings} and \cite{Anholm}). This is so because it is assumed
that stochastic gravitational waves prevail and that a single event leading to the  direct observable presented 
here would normally be too small to be observed directly. Unfortunately, the averaged pulsar-pulsar correlator 
when supplemented with the uncorrelation hypothesis normally used, likely turns it into a local observable. 
Local observables are not affected by modifications in the wave number. One needs to be able to measure the gravitational 
waves in two widely separated points for the wave number to be relevant. 
Otherwise only the frequency of the gravitational waves is relevant and as we have seen this is not at all 
varied with respect to the usual treatment. It is also for this reason that this effect is totally invisible in an 
interferometric experiment such as LIGO.

It is thus essential to validate the effect observationally to consider single events observations. Particularly in PTA,
rather than looking for stochastic gravitational-wave background. It is true that the latter may dominate, but
the effect reported here brings such a relevant enhancement that the search technique may need to be changed.
In the following sections we will provide a detailed characterization of the signal as well as its dependence on various
parameters.

\section{Characterization of the signal}
\label{SectionCharacterization}
Let us recall the controversy presented in the introduction about the actual value of the 
Hubble constant. Such apparent tension implies that the discrepancy is not method, team 
or source dependent, so a deep understanding is needed. The relevant value for $H_0$ 
in the present discussion would be the local one, as we are dealing with sources 
at $z\ll 1$. In fact, if lucky, this could provide an 
alternative method for measuring $H_0$ locally, on top of making PTA observations a lot easier.

\begin{figure}[!t]\begin{center}
	\includegraphics[scale=1.3]{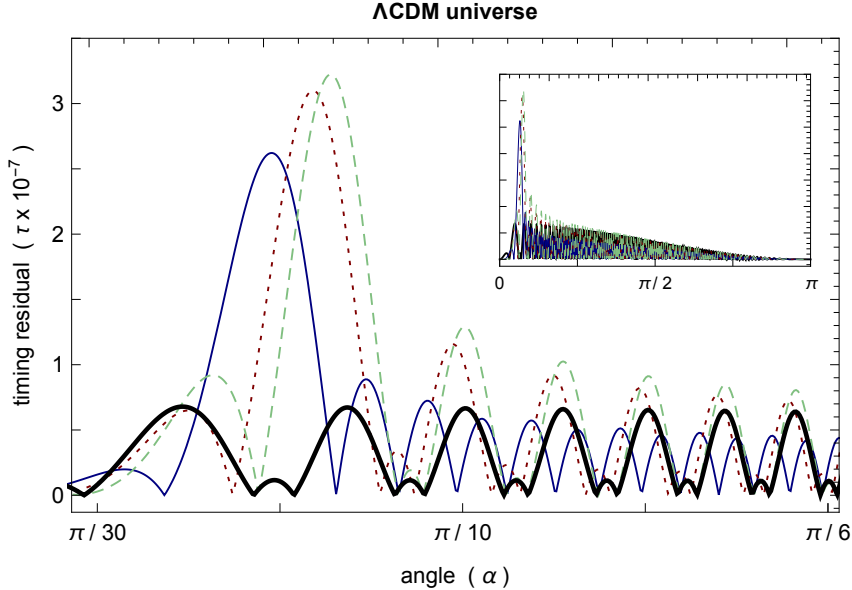}\quad
	\captionsetup{width=0.85\textwidth}	\caption{Absolute timing residual $|\tau_{ \scriptscriptstyle GW}|$ for three different values of the Hubble constant: $H_0 = 73.24 \pm 1.74$ km s$^{-1}$ Mpc$^{-1}$ reported in \cite{Riess} (green dashed curve), $H_0 = 67.4 \pm 0.5$ km s$^{-1}$ Mpc$^{-1}$ from \cite{H0Plank} (red dotted curve) and $H_0=60 \pm 6$ km s$^{-1}$ Mpc$^{-1}$ \cite{Tammann} (blue curve). We also include the propagation in Minkowski in thick black line. The remaining cosmological parameters employed for the integration do not vary from the ones used in the computation performed in Figure \ref{FigTimingResidual1Comp}. The small plot contains the four signals shown in their whole angular range for a comparison.} \label{FigTimingResidualH0}\end{center}\end{figure}
Now we would like to make a more detailed characterization of the possible signal. In Figure \ref{FigTimingResidualH0} 
we plot the modulus of the timing residual \eqref{TimingResidualH0} for three different values of the 
Hubble constant. In accordance with the discussion regarding Eq. \eqref{RelationAngleHubble}, the angular position for the maximum $\alpha_{max}$ is significantly dependent of the Hubble constant. 
In fact, an analogous relationship has been discussed in \cite{Espriu1} when gravitational waves propagate 
in a purely de Sitter universe (i.e. only $\Lambda$). Taking a fixed distance to the source $Z_E$, the authors verify in this way
the validity of the  approximate analytical expression relating the angular position of the maximum and the (local) value of the cosmological constant, i.e. $\Lambda(\alpha)$. Note that in the previous discussion we freely chose the position
of the source in the sky with respect to the chosen pulsar with the purpose of revealing the enhancement. In a real observation
the problem is of course reversed.

\subsection{Dependence on the distance to the source}
Obviously, for a fixed value of the Hubble constant, that in the following we assume to be $H_0 = 73.24 \pm 1.74$ km s$^{-1}$ Mpc$^{-1}$, we can test the dependence
of the effect on the distance to the source, and also compare the results to Eq. (\ref{RelationAngleHubble}).
In Table \ref{TableVaryZ} we have tested this relation for a range of validity of the source location in the 
corresponding approximation.
\begin{table}[!t]
    \centering
	\begin{tabular}{|c|c|c|c|c|c|c|c|c|}\hline
	\ Source distance\ &$\hspace{15pt}$Strength$\hspace{15pt}$&$\hspace{20pt}\alpha_{graphical}\hspace{20pt}$&$\hspace{5pt}\alpha_{theoretical}\hspace{5pt}$&$\hspace{10pt}$Deg$\hspace{10pt}$\\\hline\hline 100 Mpc& $1.71\times10^{-6}$&0.2125 / 0.2130&0.2101&$\sim12^{\circ}$\\500 Mpc&$1.60\times10^{-6}$& 0.4716 / 0.4720& 0.4735&$\sim27^{\circ}$\\1 Gpc&$1.67\times 10^{-6}$&0.6748 / 0.6751 &0.6761&$\sim39^{\circ}$\\\hline\end{tabular} 
	\caption{Comparison of the angular location of the maximum where the pulsar J2033$+$1734 would experience the enhancement on its timing residual. We compare the angle $\alpha_{max}$ computed in two different ways, graphically by means of numerical integrations as the one performed in Figure \ref{FigTimingResidualH0} and with the formula \eqref{RelationAngleHubble}. As can be seen the accuracy is quite remarkable.}\label{TableVaryZ}
\end{table}
We have selected the pulsar J2033$+$1734 located at $L\sim2$  kpc. Leaving the pulsar fixed, we vary 
the distance to the source for three different values of $Z_E$. We compare the angle of the peak obtained from analogous 
plots as the one presented in Figure \ref{FigTimingResidualH0}, called $\alpha_{graphical}$, and the same angle 
computed by the approximate formula \eqref{RelationAngleHubble}, $\alpha_{theoretical}$. There is a nice agreement 
between both ways of finding the angular location of the maximum. These results have been checked 
independently in \cite{Alfaro}.

\subsection{Dependence on the frequency}
From the previous analysis we have already seen that the maximum turns out to be moderately stable under changes of the parameters involved in the gravitational wave production.
\begin{figure}[!h]\begin{center}
	\includegraphics[scale=0.85]{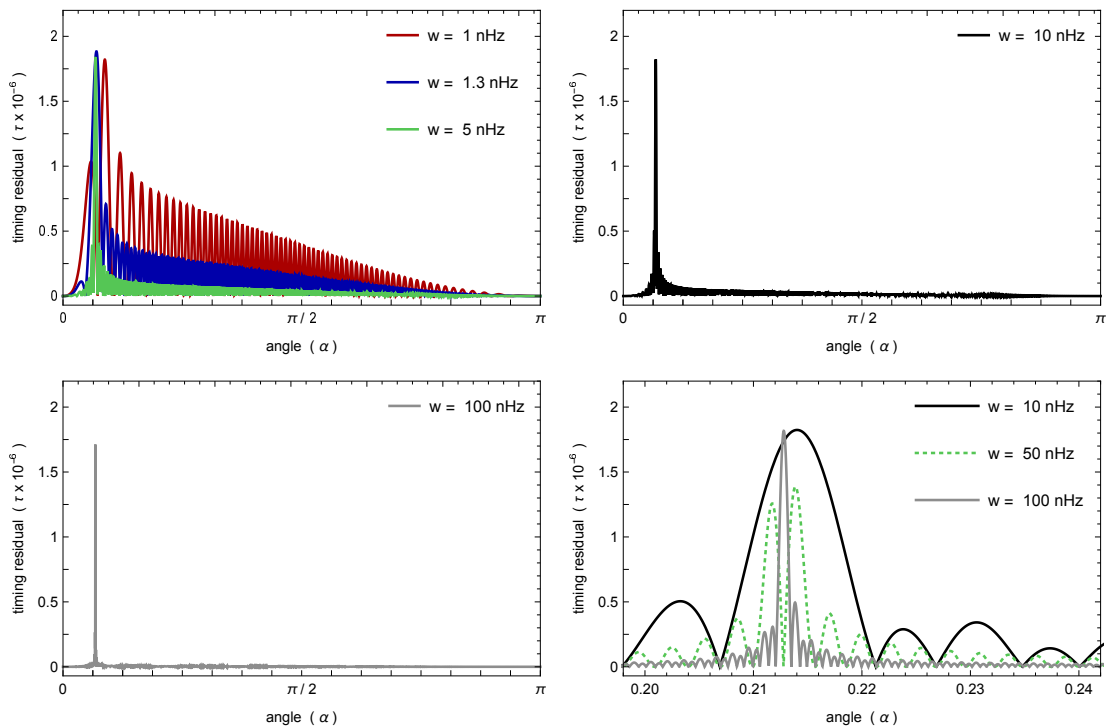}
	\caption{All curves show monochromatic gravitational waves. The first graph shows three nearby frequencies. We can see that the signal decreases quite fast its main value when the frequency is increased in some decimals. On the contrary, the peak remains nearly at the same angular position. In the second and third graphs we increase the frequency in two orders of magnitude. The signal decreases even more, while the peak does not. In the last plot the zoom makes clear that signal stretches when the frequency is increased. The same compression suffers the peak.}
	\label{Monochromatic}\end{center}\end{figure}
Let us now show the dependence of the signal when the frequency of the gravitational wave changes in the range of the characteristic values where the PTA experiment is expected to be performing. In Figure \ref{Monochromatic} the absolute value of the timing residuals for the angular coordinate, $|\tau_{\scriptscriptstyle GW}(\alpha)|$, is plotted  when the frequency increases. This magnitude would be registered by the pulsar J2033$+$1734 located at $2$ kpc.

In the first graph we superpose three different harmonic functions with slightly different values for the frequency, $w=1$ nHz, $w=1.3$ nHz and $w=5$ nHz. For values near  $\sim 1$ nHz, the peak already doubles the rest of the signal. If we slightly increase the frequency by a few decimals, the signal falls while the peak remains stable with nearly the same height. We chose these values to show how fast this decay results with changes in the frequency: comparing the signal computed with $w=1$ nHz with the one obtained for $w=1,3$ nHz), the signal falls by half, while the peak does not change noticeably in height or location. This pattern remains valid for all frequencies tested. At $w= 5$ nHz the peak is already one order of magnitude higher than the rest of the signal. It is also noticeable that the peak becomes narrower as the frequency increases.

The fourth plot shows a zoom on the enhancement region for the latter two figures. Here again it is clear how the signal, peak and oscillations, stretches. The signal also falls while the peak remains at the same place with the same height. Sometimes two peaks next to each other appear, however, as can be seen in the signal computed with $w=50$ nHz (green dotted curve), the value of the maximum is still approximately one order of magnitude above the signal. In conclusion, the height of the signal turns out to be nearly constant as the frequency of the gravitational wave varies.

\begin{figure}[!b]\begin{center}
	\includegraphics[scale=0.8]{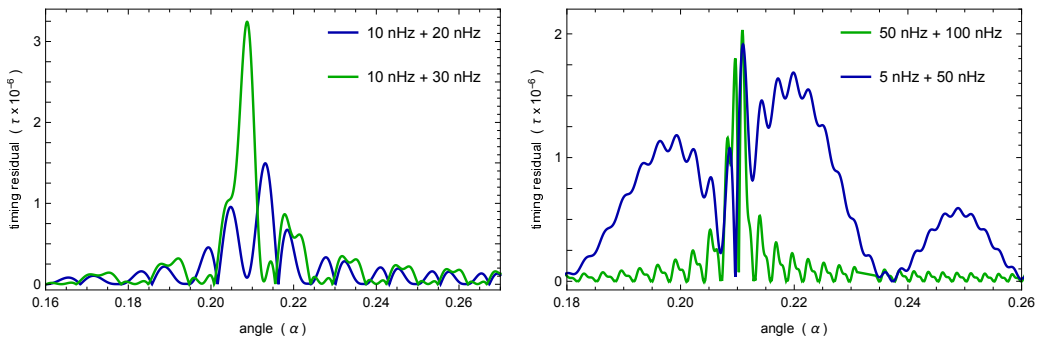}
	\caption{Absolute timing residual $|\tau_{\scriptscriptstyle GW}(\alpha)|$ for non-monochromatic gravitational waves. The first plot combines two frequency with nearer values. It is seen that the peak suffers some modifications in its height. In the second plot, the frequencies differ in an order of magnitude and the enhancement zone becomes wider. The principal peak remains normally in the same position. }\label{NonMonochromatic}\end{center}\end{figure}

It is interesting to explore signals consisting in the superposition of several frequencies. In Figure 
\ref{NonMonochromatic} the absolute value of the timing residual is computed employing bichromatic waves. 
The main characteristics of the signal remain the same if we plot for the full angular range $\alpha\in[0,\pi]$. 
We can see an oscillating signal around zero and a remarkable peak always at the same location. 
The peak is seen to depend only on the value of the Hubble constant 
and on the distance to the source. A closer look reveals that the peak suffers several deformations when 
two or more frequencies are taken into account, but it is always present. In this figure we show two combinations
of two frequencies each; one combines relatively similar frequencies while the other makes a superposition
of rather different ones. 
In summary, the main peak of the signal remains at the same location when frequencies are varied or combined. However, the
combination of frequencies has a desirable effect as it enlarges the range of angles where the enhancement is visible, and this 
is quite welcome when it comes to a possible detection of the effect.

\subsection{Time dependent amplitudes}
Let us assume a varying global factor for the wavefront such that
\begin{equation}
	d\tau_{\scriptscriptstyle GW}\,\rightarrow\,A(T)\,d\tau_{\scriptscriptstyle GW}.
\end{equation}
$d\tau_{\scriptscriptstyle GW}$ is the differential timing residual defined as the integrand in Eq. \eqref{TimingResidualH0}. 
The next step is to give a structure to the form in which the amplitude varies. Let us impose two different 
behaviours: a signal modified by a very slowly varying exponential and by means of a linear varying amplitude.
For the former, we would have
\begin{equation}\label{Amplitude}
	A(T)=\exp\left\{\frac{c}{L}\ \left[T-\Bigl(\ T_E-\frac{L}{c}\ \Bigr)\right] \right\}=\exp\left(1+x\right)\ .\end{equation}
For the second equality we have used the parametrization of the Earth-pulsar path. Then we perform the numerical 
integration as in \eqref{TimingResidualH0}, with the argument of the trigonometric function given by \eqref{Theta}, 
but taking into account the new varying amplitude. 
In such manner, when the gravitational wave arrives to the pulsar location, the global factor in \eqref{Amplitude} equals $1$ (the integration starts at $x=-1$), then it grows while the integration is performed. 
At the endpoint, the global factor reaches a value equal to $A(T_E)=e$. While this may not be a realistic profile of the wave front, it serves us to test the dependence on $A(T)$.

\begin{figure}[!b]\begin{center}
	\includegraphics[scale=0.81]{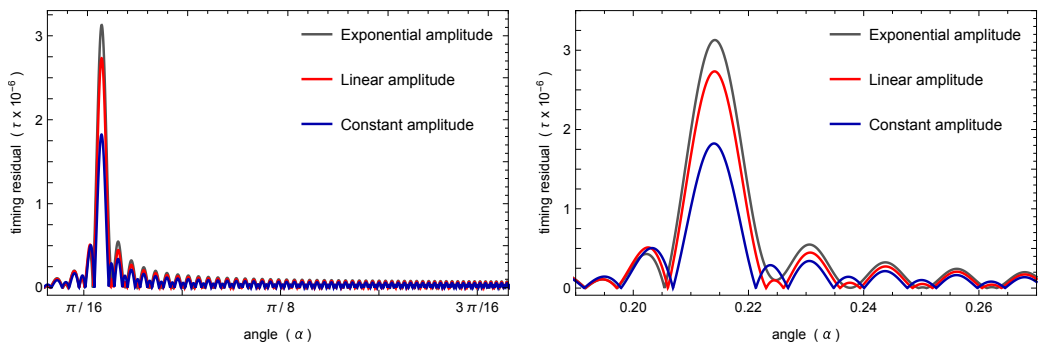}
	\captionsetup{width=0.85\textwidth}
	\caption{Absolute timing residual $|\tau_{\scriptscriptstyle GW}(\alpha)|$ for gravitational waves characterized by the two different time varying amplitudes mentioned. As a comparison we also include the timing residual corresponding to a unitary amplitude in blue.}\label{AmplitudeVary}\end{center}\end{figure}

The same steps have been done for a linearly varying amplitude. We start with a linear structure $A(T)=A_0(T-T_0+b)$ and impose the same boundary conditions. At the start of integration, we want that the amplitude takes a unit value; for this to happen we must impose $b=1/A_0$. We have used again the same parametrization and we take $A_0=c/L$. With these considerations, we use 
\begin{equation}
A(T)=1+(1+x) \end{equation}
to compute the timing residual \eqref{TimingResidualH0}.

In Figure \ref{AmplitudeVary} we show how the timing residual changes when the above time-dependent amplitudes are introduced. One more time we take the signals coming from the pulsar J2033$+$1734. The timing dependent amplitudes modify the overall characteristics in the expected way, increasing the whole signal according to the function taken. 

\subsection{Dependence on the distance to the pulsars}
\begin{figure}[!b]\begin{center}
	\includegraphics[scale=1.2]{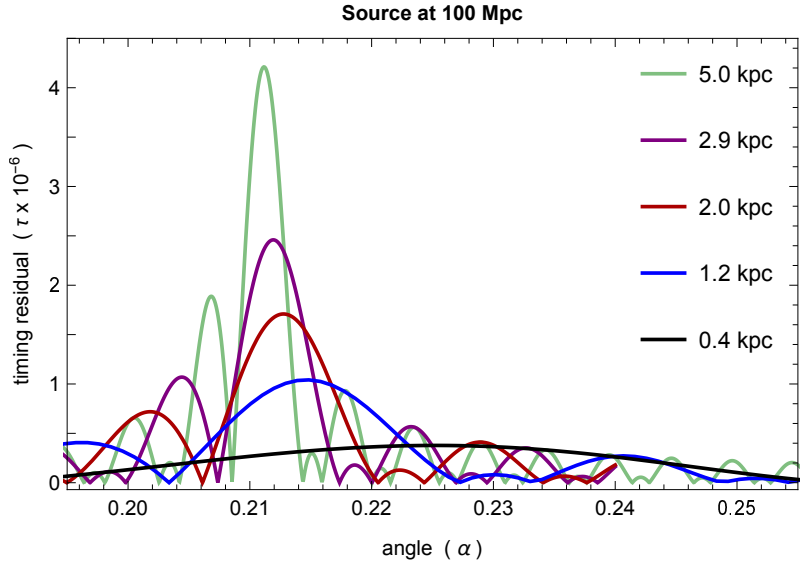}
	\caption{Zoom of $|\tau_{\scriptscriptstyle GW}|$ on the enhancement region. Each curve represents each of the 5 pulsars chosen with different distances to the Earth. It is worth to note that all peaks are way higher than the rest of the signal. Even the lowest in height, J0751$+$1807 (in black), is significantly bigger (a factor 5) than the rest of the remaining signal. For instance, take as an example the blue curve corresponding to the pulsar J1603$-$7202; this pulsar has nearly the distance employed in Figure \ref{FigTimingResidualH0} (where for the red curve $L$ was fixed to $1$ kpc). The remaining pulsars are: J2033$+$1734 in red, J1802$-$2124 in violet and J1939$+$2134 in green.} \label{FigLdependence}\end{center}\end{figure}

Let us focus on the dependence of the peak on the distance $L$ to the pulsar. Let us fix the gravitational waves source at $100$ Mpc. Then, we plot the absolute value for the timing residual 
versus the angle subtended by the source and the measured pulsar as seen from the observer, $|\tau_{\scriptscriptstyle GW}(\alpha)|$, for different millisecond pulsars from Table \ref{TablePulsars}. 
We choose five pulsars with halfway apart locations with respect to the Earth; they are J1939$+$2134, J1802$-$2124, J2033$+$1734, J1603$-$7202 and J0751$+$1807 in a decreasing order of distance from us. 
In Figure \ref{FigLdependence} we show a zoom of the enhancement region.
As expected, as long as the Hubble parameter and the distance to the source are fixed parameters, the angular position 
where the enhancement peak is located do not vary for the different curves. On the contrary, the 
strength of the peak grows to the extent that farther away pulsars are used in the calculation. There is a logic
behind this, as the accumulative effect of the enhancement becomes larger as it does the optical path.

We emphasize that in all cases the enhancement peak is at least an order of magnitude larger than the rest of the signal. 
Even for the black curve in Figure \ref{FigLdependence} where at first sight and because of the scale seems 
that there is no peak, it reappears when one plots for a wide range of values for $\alpha$. Note that 
for this pulsar, J0751$+$1807, $L=0.4$ kpc which is a characteristic distance very similar to the one used 
in Figure \ref{FigTimingResidualH0} where $L=0.32$ kpc. In fact its corresponding plot is very similar 
to those curves too, where it is seen that the peak is still significantly bigger than the whole signal.

What we also see from these figures is that while the peak becomes higher when the distance to the pulsar grows, it
becomes also narrower. So peaks resulting from closer distances are broader. 
Figure \ref{FigLdependence} indicates that pulsars located in a region 
characterized by $\overline\alpha\in[\alpha_1;\alpha_2]$ register an anomalous enhancement in their timing residual because of the passage of a gravitational wave produced at a distant source. This interval is related with the width of the peak; in fact the global maximum is the center of this interval: $\alpha_{max}=(\alpha_1+\alpha_2)/2$. All in all, pulsars placed more near to us will have a bigger range of values $\overline\alpha$ where $\tau_{\scriptscriptstyle GW}$ would register a significant enhancement.
We note that, as reported in the first data release of the IPTA collaboration in \cite{Verbiest}, almost the 25 \% of the pulsars considered are at distances $L \leq 0.5$ kpc from us.

\begin{table}[!t]\centering\begin{tabular}{|l|r|l|}\hline
	Pulsar&distance (kpc)&C\\\hline\hline J0030$+$0451& \begin{minipage}{0.15\linewidth}\vspace{0.5pt}$0.28^{+0.10}_{-0.06}$ kpc\end{minipage}&N,E\\J0610$-$2100& 3.5$\pm$0.7 kpc &E\\ J0751$+$1807&$0.4^{+0.2}_{-0.1}$ kpc&E\\J1603$-$7202&$1.2\pm 0.2$ kpc&P\\J1802$-$2124&$2.9\pm0.6$ kpc&E\\\hline \end{tabular}\hspace{20pt}\begin{tabular}{|l|r|l|}\hline
	Pulsar&distance (kpc)&C\\\hline\hline J1804$-$2717&$0.8\pm0.2$ kpc&E\\J2033$+$1734&$2\pm0.4$ kpc&E\\J1939$+$2134&$5^{+2} _{-1}$ kpc &E,P\\J1955$+$2908&$4.6\pm0.9$ kpc&N,E\\&&\\\hline \end{tabular}
	\caption{Pulsars employed in the analysis \cite{ATNF}. The distances to the pulsars have been taken from the reported values in \cite{Verbiest}. The column C referees to the collaboration monitoring the pulsar: N stands for the NanoGRAV collaboration, E for the EPTA and P for PPTA.}\label{TablePulsars}\end{table}
	
\begin{figure}[!b]\begin{center}
	\includegraphics[scale=0.95]{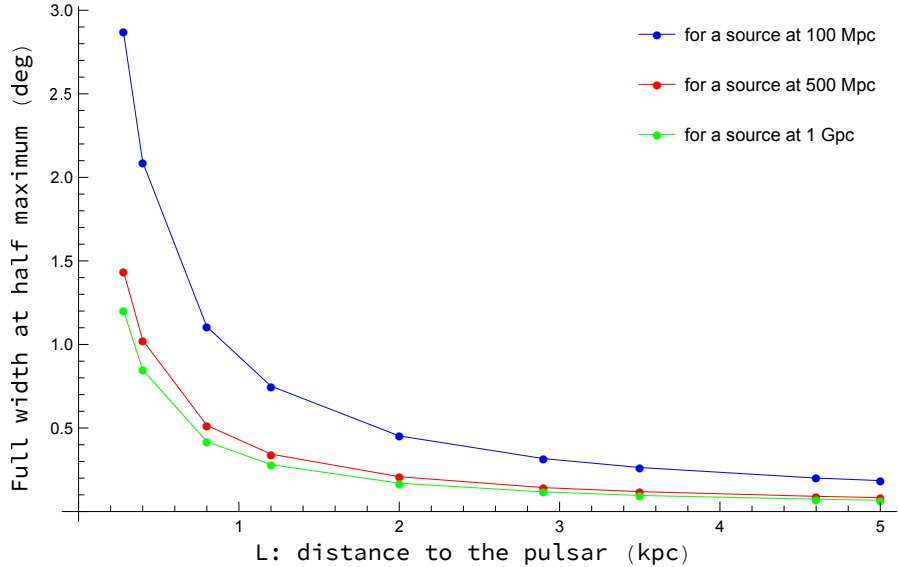}\vspace{5pt}
	\caption{Each point represents the angular value using the FWHM criterion for obtaining the main peak width of the signal. We compute this magnitude for the distance to each of the 9 pulsars in Table \ref{TablePulsars}, order in the $x$-axis by the distance from the pulsar to the Earth. Each curve corresponds as indicated to a different source location.}\label{FigWithVsL} \end{center}
\end{figure}

As a more realistic example we will make use of all the real pulsars listed in Table \ref{TablePulsars}. Using three different source locations, we can draw the $\alpha$ profile of the modulus of the timing residuals corresponding to a fixed distance. Added to this, we need a criterion to link the width of the peak with the interval $\overline{\alpha}$ where the enhancement happens. A conservative election is the {\it full width at half maximum} (FWHM) parameter. It is defined as the distance between points on the curve at which the function reaches half of its maximum value\footnote{Greater distances from Earth would register a higher but narrower peak. The FWHM measure can be flexibilized; the half peak is quite a big threshold that can be reduced significantly in pursuit of increasing the width where the enhancement occurs. This would significantly increase the areas where this effect can be detected in the sky.}. In Figure \ref{FigWithVsL} we show the result of computing the FWHM for the distance to the 9 pulsars shown in Table \ref{TablePulsars}, when the gravitational wave source is fixed arbitrarily at 3 different distances (100 Mpc, 500 Mpc and 1 Gpc). 

\begin{figure}[t!]\begin{center}
	\includegraphics[scale=0.75]{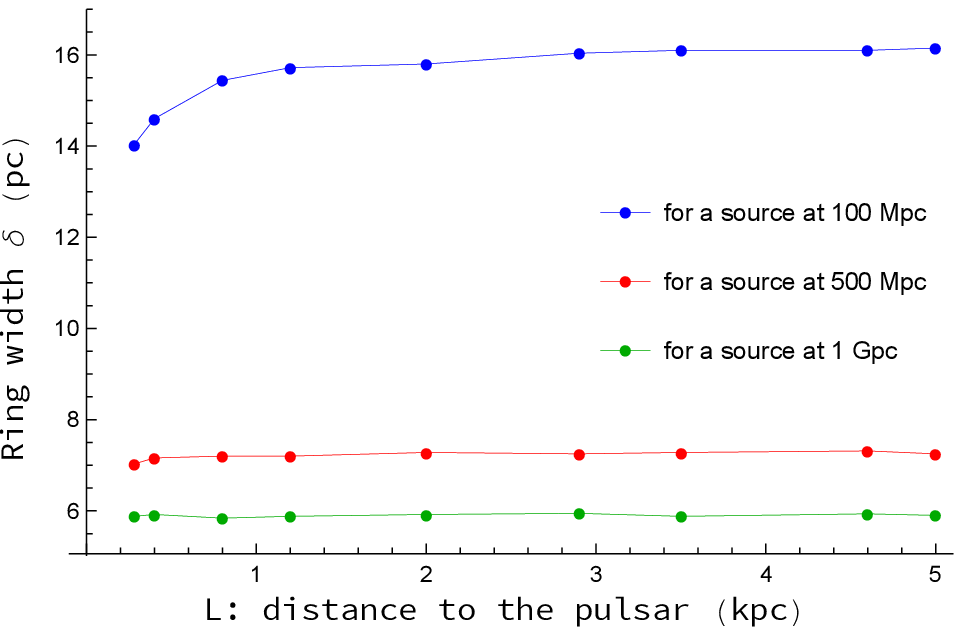}\hspace{39pt}\includegraphics[scale=0.4]{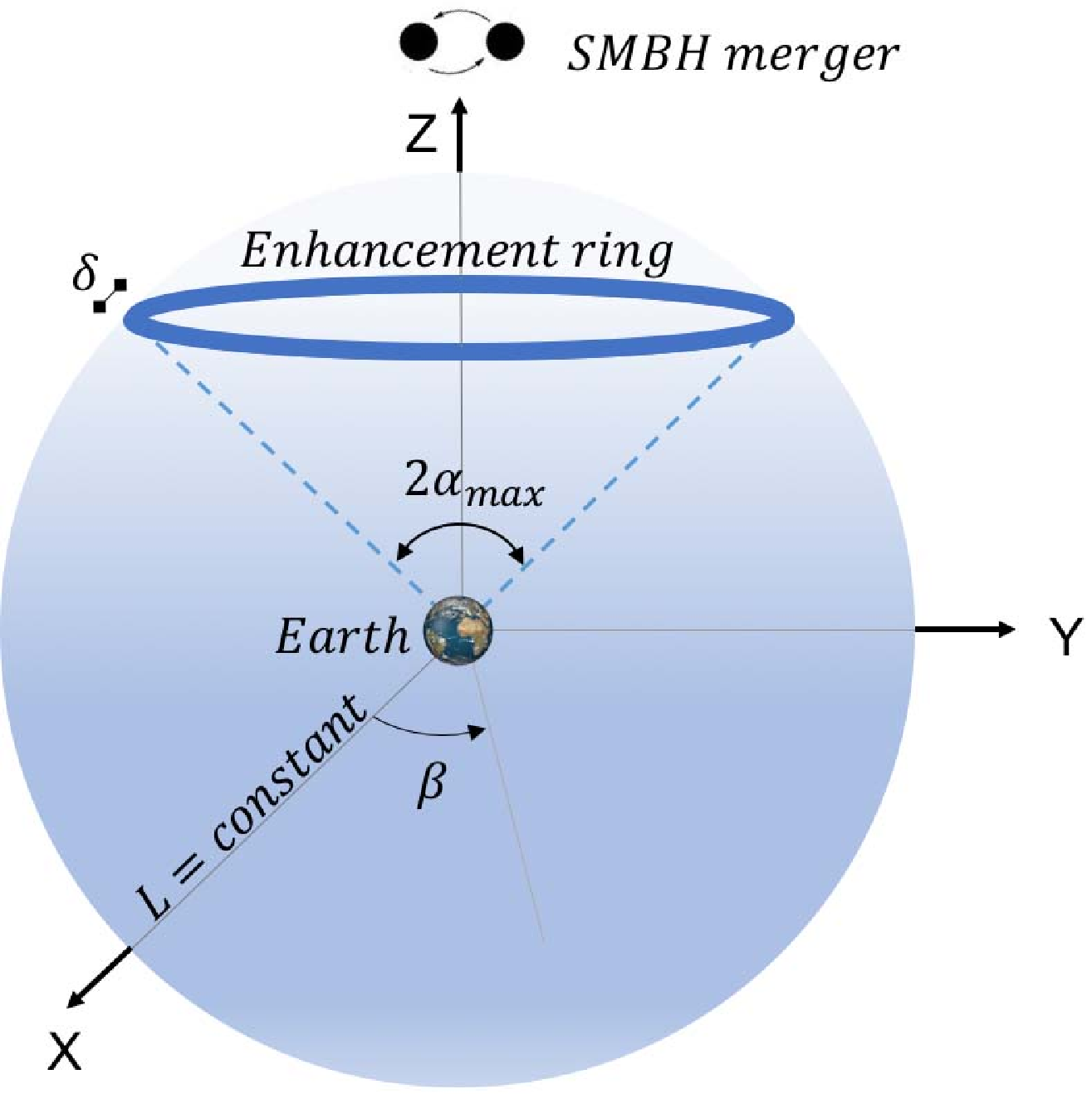}\vspace{5pt}
	\caption{Spatial region (in blue) where any falling pulsar would experience an anomalous enhancement in timing residual $\tau_{\scriptscriptstyle GW}$. This region is shown for only one value of the distance $L$, but appears for all values of $L$ with nearly the same characteristics (the complete region is a truncated cone-like surface). On the left we show the width of the ring that seems to depend only on changes of the source location $Z_E$.}\label{FigRing}\end{center}\end{figure}

We can convert the angular interval where the enhancement takes place to astronomical distances; 
i.e. $\overline{\alpha}\rightarrow \delta $. Taking into account at which distance the pulsar 
would be located, we can use trigonometry to calculate the size in parsecs that would correspond to a determined angular 
interval subtended at that distance $L$. For the distances to the 9 pulsars referred previously, we obtain the plot in Figure \ref{FigRing}. 
A source located at the referred distance would ``light up" a zone characterized by an angular 
width $\overline{\alpha}$. This interval would correspond to a distance $\delta$ as it is shown in the 
plot on the left. Any pulsar falling in this interval would experience a significant enhancement in its timing residual.
The three different curves show that the size of this zone is of a few tens of parsecs and its width is nearly constant 
with respect to the distance $L$ (the fluctuations of each curve when varying the location $L$ of the pulsar represent 
more or less 10\% of its value). We can conclude that the width of the peak seems to be highly dependent on the location 
of the gravitational wave source $Z_E$, as it was for its position $\alpha_{max}$ derived in Eq. \eqref{RelationAngleHubble}.

There is another important aspect to take into account. According to the definition of the timing residual, the effect is $\beta$ independent and nearly $L$ independent. The $\beta$ independency of the theory is translated into a symmetry of revolution for the enhancement region $\delta$ around the $Z$-axis (the axis coinciding with the direction to the source). This means that a supermassive black hole merger would generate a ring of enhancement at each constant distance $L$ from the Earth characterized by the azimuthal angle ranging between $\beta\in[0;2\pi)$. The angular radius of the ring is given by the polar angle $\alpha_{max}$ (recall the results on Table \ref{TableVaryZ} where different angular locations of the enhancement have been obtained by using different distances to the source) and its width $\delta$ is given in parsecs on the plot next to the scheme of Figure \ref{FigRing}, according to the corresponding distance involved. This region where a significant deviation in the timing residual would be registered is shown in the scheme as the blue ring and will be called the {\sl enhancement region}. The invariance with respect to $L$ makes this region a truncated cone-like surface more than a thick ring; however for practical issues that will be discussed in the next section, let us retain the region as a ring-like image such that pulsars placed there would be seen by us as experiencing a notorious deviation on the time of arrival of their pulses. The width of this ring-like enhancement region will be in the 1 to 10 pc range, depending on the distance to the source.

\section{The Pulsar Timing Array project}
The quest for the detection of gravitational waves in PTA follows two main avenues. One proposal focuses on 
stochastic (isotropic and incoherent) gravitational-wave background \cite{Hellings,FosterBacker1990Kaspi1994,Jenet2005}
produced by incoherent superposition of radiation from the whole cosmic population. In \cite{Verbiest} a brief review 
goes through the main sources for this background and define the resulting signal that would (or not) be detected by PTA.

More interesting for our purposes is the search for single or individual sources, if they are sufficiently bright. 
Clearly, a nearby binary merger would be detectable if and only if the gravitational signal stand out above the 
root-mean-square value of aforementioned background \cite{Sesanna2009}.
There is controversy on whether a single merger event is detectable by PTA 
\cite{Seto2009vanHaasteren2010Pshirkov2010Favatara2009}.
This debate and its somewhat pessimistic conclusion assumes of course the standard analysis of the signal-to-noise ratio.

As of today there is no conclusive evidence of an indirect measurement of any gravitational radiation in this regime 
of frequencies wherever they are produced. Single-source limits show that all proposed binary mergers are still 
below current sensitivity levels: recent limits have been derived in \cite{Babak2015} from the EPTA, 
in \cite{Arzoumanian2014}  from the NANOGrav data and in \cite{Zhu2014} from the PPTA collaboration. In addition, 
similar conclusions have been reached for gravitational wave bursts \cite{Wang}. Our analysis reported here may 
suggest reconsidering the expected signal and how to look for it.

When gravitational waves propagate over flat spacetime the phase effect discussed before is lost as we have shown 
in the first plot of Figure \ref{FigTimingResidual1Comp} (Minkowskian propagation). Including the  
familiar redshift in frequency alone as the only way 
of including the effect of the FLRW metric, without taking in consideration the effect discussed in this work,
for each single-source event a ring-like region of enhancement in the sky is possibly missed. The enhancement should be 
affecting any pulsars in it, irrespective of the distance to the observer $L$. We recall that the enhancement may
represent an increase in the signal of more than one order of magnitude. From the study presented in the previous sections
and the typical values of $L$ we concluded that the width of this ring is of the order of 1 to 10 parsec.

\begin{figure}[t!]\begin{center}
	\includegraphics[scale=0.7]{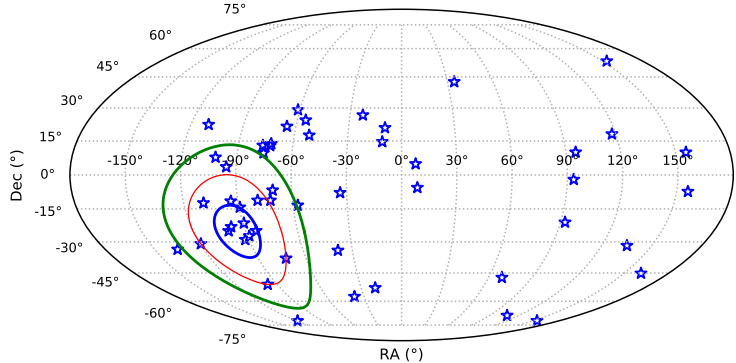}\vspace{-40pt}
	\caption{Galactic distribution of all MSPs observed by the IPTA are indicated by blue stars. Enhancement rings correspond to three different sources placed at 100 Mpc in blue and at 1 Gpc in green. The red ellipse would be the enhancement ring if the three further south pulsars are supposed to perceive the enhancement. When we see that at least three pulsars register an anomalous timing residual, we have enough information to know where the effect has its origin. These three points determine an enhancement ring of $\alpha_{max}\simeq25^\circ$ with respect to a center where there should be placed a hypothetical gravitational wave source. According to the relation \eqref{RelationAngleHubble}, the angle of enhancement constrains the spatial location of the source to $Z_E\simeq443$ Mpc. The angular radius subtended between the source and the pulsar are for the other two enhancement rings: $\sim39^\circ$ for the green ellipse and $\sim12^\circ$ for the blue one.}\label{FigMollwide} \end{center}\end{figure}

The image presented in Figure \ref{FigRing} reminds us of the coordinates of the pulsars monitored by the IPTA
collaboration, irrespective of the value of $L$, the distance between the pulsar and the Earth. 
In the graphical representation on Figure \ref{FigMollwide}, for each sphere of the sky at a distance $L$, a  
supermassive black hole merger would induce a circular ring-like region with a diameter given by twice the angle 
subtended between the source and the pulsar with respect to the source. In Figure \ref{FigMollwide} we show in a 
Mollwide projection the location of these pulsars together with the enhancement ring for a source located at 100 Mpc and at 1 Gpc.

Assuming that each collaboration measures the timing residual over an adequate integration 
period a single-event should represent a simultaneous firing of every pulsar laying inside the enhancement zone. If the collaboration finds at least three pulsars experiencing an enhancement in their timing residuals, each of such triple signals would define a vector pointing towards the source. Once this direction is tentatively determined, 
a more detailed analysis of the signal's effect on other pulsars would be possible.

Note that in the previous section all signals refer to snapshots at a given specified time. Of course, there is a large 
background that obscures the signal (even taking into account the enhancement) and therefore the signal has to be integrated
over an adequate period of time. This issue has not been considered here, but we refer the reader to the work \cite{Espriu1} where
it was preliminarily estimated that observation of the given set of pulsars over a period of three years with an eleven-day
period would provide a sufficiently good signal. Needless to say that this is a very crude value that needs careful consideration.

\newpage
\section{Conclusions}
The work reported in this article is the continuation of previous studies \cite{Espriu1,Espriu2,Espriu3,Espriu4}. First of all, we have confirmed
the validity of the approximate solution shown in Eq. \eqref{gwfrw} by solving directly, up to ${\cal O}(H_0)$, the wave equation for linearized perturbations in a FLRW geometry. This approximate solution is valid to describe gravitational waves
originating from sources up to $\sim$ 1 Gpc or, equivalently, up to a redshift of 0.2 approximately. Beyond that, higher
order corrections are necessary and these are being considered at present.

It is important to emphasize that, at this order, all cosmological parameters enter in the form of the Hubble constant 
$H_0$, but it is not so when higher orders are included.

The solution has an effective wave number $k_{eff} \neq w_{eff}$, where $w_{eff}$ is the familiar redshifted frequency. This
does not mean that gravitational waves propagate subluminically; they do so when the proper `ruler' distance is considered. 
But it has a definite impact in observations of gravitational waves involving non-local effects, and there is an
integration over a path in comoving distances. This is the case of observations made in the framework of the IPTA project 
(and, incidentally, also in LISA, even though the analysis in this case is yet to be performed).

We have discussed in great detail the dependence of the effect on various parameters influencing the observation: 
superposition of frequencies, time-varying amplitudes, distance to the source, distance to the intervening pulsars. In all
cases the signal should be perfectly visible, well above the usual analysis assumed. The analysis presented hereby
may help to establish a well-designed search strategy. We note that one aspect we have not discussed at all is the 
integration times. In \cite{Espriu1,Espriu2} an observational strategy was mentioned, but we consider that the collaboration between the different consortia is crucial in this point.

In conclusion, the effect reported here appears to be firmly established, it may provide an opportunity for the IPTA 
collaboration to detect correlated signals in triplets of pulsars and, from there, determine the location and other properties
of the source of gravitational waves. Needless to say that an independent local determination of $H_0$ (and eventually $\Lambda$) 
would be of enormous interest.

\section*{Acknowledgements}
Funding for this work was provided by the Spanish MCIU under project MDM-2014-0369 of ICCUB 
(Unidad de Excelencia `Maria de Maeztu'), grant FPA2016-76005-C2-1-P and grant 2017SGR0929 (Generalitat de Catalunya). L.G.
acknowledges the receipt of FPI grant BES-2014-067939.
We are grateful to Jorge Alfaro, Mauricio Gamonal and Santiago del Palacio for interesting discussions and 
collaboration in various aspects of this work. We also 
thank J. Bernabeu, D. Puigdom\`enech and I. Saiz de Lucas.

\end{document}